\documentclass[reprint,amsmath,amssymb,aps,prb,floatfix]{revtex4-2}

\usepackage{graphicx}
\usepackage{dcolumn}
\usepackage{bm}
\usepackage{float}

\usepackage{mystyle}
\usepackage{braket}

\begin{document}

\preprint{APS/123-QED}

\title{Spin-triplet paired Wigner crystal stabilized by quantum geometry}
\author{Dmitry Zverevich}
\author{Alex Levchenko}
\author{Ilya Esterlis}
\affiliation{Department of Physics, University of Wisconsin-Madison, Madison, Wisconsin 53706, USA}

\date{\today}

\begin{abstract} 
We have used variational states to analyze the effects of band geometry on the two-dimensional Wigner crystal with one and two electrons per unit cell. At sufficiently low electron densities, we find that increasing Berry curvature drives a transition into a crystalline state composed of spin-triplet pairs carrying relative orbital angular momentum $m=-1$. The essential features of this transition are captured by an effective two-electron quantum dot problem in the presence of Berry curvature. Our results point to a purely electronic, strong-coupling mechanism for local spin-triplet pairing in correlated two-dimensional electron systems with quantum geometry.
\end{abstract}

\maketitle

At sufficiently low densities, a clean electron gas crystallizes into a Wigner crystal (WC) due to the dominance of Coulomb interaction energy over kinetic energy \cite{wigner1934}. In the extreme dilute limit, semiclassical considerations predict that the electrons form a Bravais lattice with one electron in each unit cell. At higher densities, however, quantum corrections become important, and some years ago Moulopoulos and Ashcroft suggested that quantum effects may stabilize \textit{non-Bravais} lattices of electrons, with \textit{electron pairs} residing at the lattice sites \cite{moulopoulos1992}. The paired WC becomes energetically favored over the ordinary WC due to a reduction of zero-point energy that can overcome the relatively small Coulomb energy cost of pairing electrons within a unit cell. Subsequent work confirmed a region of stability for the paired WC \cite{moulopoulos1993,taut1994,taut2001}, albeit in a density range where the ground state is expected to be a homogeneous liquid \cite{tanatar1989,attaccalite2002,drummond2009,azadi2024,smith2024}. Nevertheless, these calculations suggest a strong tendency toward local pairing of electrons over a broad density range in the strongly correlated regime.

Here, we investigate paired electron crystals in electronic bands with Berry curvature and quantum geometry. Our study is motivated by the widespread observation of electron crystals in two-dimensional graphene and transition-metal dichalcogenide van der Waals materials \cite{smolenski2021,sung2025, zhou2021,xiang2025, tsui2024, lu2024, lu2025, walters2025, su2025,tan2025ideal}, and the interplay between band geometry and strong correlations in these systems, which has attracted substantial recent theoretical attention \cite{joy2023,zheng2024sublattice,zhou2024,tan2024,zdong2024,patri2024,dong2024,soejima2024,dong2024stability,tan2025,bzhou2025,soejima2025,desrochers2025,desrochers20252,bernevig2025,joy2025,kim2025,zeng2025,valenti2025} (see Ref.~\cite{zhou2025} for a recent review). Of particular interest is rhombohedral multilayer graphene, where tuning the substrate alignment drives the system from a putative topological electron crystal (the ``anomalous Hall crystal") \cite{lu2024,lu2025} into a host of exotic spin- and valley-polarized superconductors \cite{han2025}, hinting at a possible connection between electron crystallization, band geometry, and superconductivity. 

Starting from an effective one-band model for a valley-polarized, spin-degenerate two-dimensional electron gas (2DEG), we show that quantum geometry stabilizes the paired crystal down to lower densities and that sufficiently strong Berry curvature drives a transition into a crystalline state of spin-triplet pairs with orbital angular momentum $m=-1$. This bears some similarity to the well-known phenomenon of singlet-triplet oscillations in two-electron quantum dots as a function of  perpendicular magnetic field \cite{wagner1992,ashoori1993}. These effects are quantum mechanical in origin, arising from the momentum-space Berry flux acquired by the electrons through zero-point motion.  We thus demonstrate a mechanism by which the combination of strong correlations and quantum geometry promotes local, spin-triplet pairing of electrons. We note recent work showing that, in Bernal bilayer graphene, sufficient Berry curvature can drive a transition to a \textit{monatomic} WC state in which electrons acquire a spontaneous orbital angular momentum \cite{joy2025}.

To model the effects of Berry curvature and quantum geometry in the valley-polarized 2DEG, we adopt the ``parent Berry curvature" model of Ref.~\cite{tan2024}, which consists of a quadratic electron band with effective mass $m^{*}$ and a continuously tunable, uniform Berry curvature $\Omega$. The Hamiltonian is given by
    \begin{widetext}
    \be
    H = \sum_{\bfk\sigma} \frac{\hbar^2k^2}{2m^*}c^\dag_{\bfk\sigma}c_{\bfk\sigma} +\frac{1}{2A}\sum_{\bfq \neq 0} v_\bfq \sum_{\substack{\bfk\bfk' \\ \sigma \sigma'}}
    F_{\bfk+\bfq,\bfk}F_{\bfk'-\bfq,\bfk'}c^\dag_{\bfk+\bfq\sigma}c^\dag_{\bfk'-\bfq\sigma'}c_{\bfk'\sigma'}c_{\bfk\sigma},
    \label{eq:H}
    \ee
    \end{widetext}
where $v_\bfq = 2\pi e^2/q$ is the Fourier transform of the Coulomb interaction, $A$ is the area of the system, and the form factors are expressed via overlaps of the Bloch functions
    \be
    F_{\bfk, \bfk'} =\braket{u_{\bfk}|u_{\bfk^\prime}}= e^{-i\frac{\Omega}{2}\bfk \times \bfk'}e^{-\frac{\Omega}{4}|\bfk - \bfk'|^2}.
    \label{eq:form_factor}
    \ee
The first factor is due to the Berry phase acquired by the electrons as they move through $k$-space, while the second term  arises due to the distance between Bloch states and is required to satisfy the trace condition \cite{roy2014}. In the limit $\Omega \to 0$, the Hamiltonian \eqref{eq:H} reduces to the conventional jellium model of the 2DEG.  

The properties of the model \eqref{eq:H} are determined by two dimensionless parameters: the first is the usual electron gas parameter $r_s$, which measures the ratio of Coulomb to kinetic energy and is conventionally defined as $r_s = a/a_B$. Here, the interparticle distance $a$ is determined by the electron density $n$ according to $n \pi a^2 = 1$, and $a_B = \hbar ^2/m^{*}e^2$ is the effective Bohr radius. The second dimensionless parameter controls the effects of quantum geometry, which we write as $\omega = \Omega/r_s^{3/2}a_B^2$; the Berry curvature $\Omega$ has dimensions of an area, and the natural unit to measure it in the present problem is set by the $k$-space extent of the localized electronic wave functions in the WC, which is of the order $\sim 1/r_s^{3/2} a_B^2$. 

\begin{figure}
    \centering
    \includegraphics[width=\linewidth]{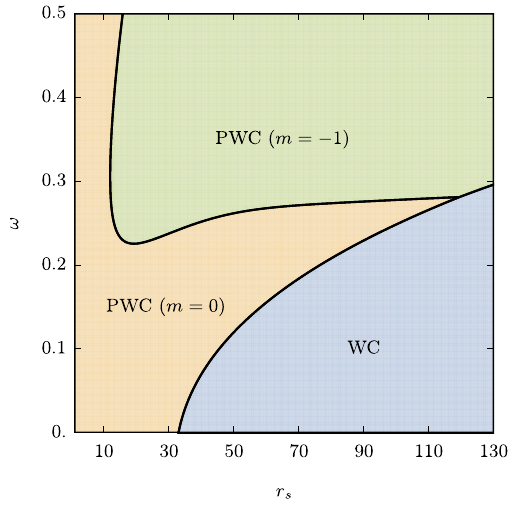}
    \caption{Variational ground state phase diagram as a function of interaction strength $r_s$ and Berry curvature $\omega$. Only crystalline phases are shown. At sufficiently large $r_s$ and $\omega$, a paired Wigner crystal (PWC) composed of orbital angular momentum $m=-1$, spin-triplet electron pairs is energetically favored. Competing crystalline phases are the monatomic Wigner crystal (WC) and PWC with $m=0$, spin-singlet pairs.}
    \label{fig:phase_diag}
\end{figure}

Our main results are summarized in the variational ground state phase diagram shown in Fig.~\ref{fig:phase_diag}. Starting with $\Omega = 0$, we find a transition at $r_s\approx 33$ from the monatomic triangular lattice WC to a triangular lattice WC composed of spin-singlet pairs with relative orbital angular momentum $m=0$ \footnote{We restrict our attention to the triangular lattice, which minimizes the classical electrostatic energy of the system \cite{bonsall1977}. We do not address the melting of the WC phases, which appears to be more subtle in the presence of Berry curvature: a simple Lindemann-criterion estimate suggests that increasing Berry curvature \textit{raises} the critical $r_s$ for melting \cite{joy2023,joy2025}, whereas recent variational Monte Carlo calculations \cite{valenti2025} find the opposite trend, with Berry curvature \textit{lowering} the critical $r_s$. We thank Brian Skinner for bringing this point to our attention.}.  
As $\Omega$ is increased, the paired WC is favored to larger $r_s$ (lower density), demonstrating that quantum geometry stabilizes local pairing deeper into the strongly correlated regime. Increasing $\Omega$ further, the paired $m=0$ crystal eventually gives way to the paired state with $m=-1$ (orbital angular momentum anti-aligned with Berry curvature). At $r_s\approx 120$, the ordinary and paired WC phases meet at a ``triple point", beyond which there is a direct transition from  the ordinary WC to the $m=-1$ paired WC. The absence of the $m=-1$ paired crystal for $r_s \lesssim 15$ implies that the emergence of pairing with nonzero orbital angular momentum arises from the cooperative effects of strong correlations and quantum geometry. 

For the ordinary WC, we have used an $N$-electron variational state composed of localized Gaussian orbitals \cite{moulopoulos1993,vandijk1991,bonev2008}:
    \be
    | \Psi \rangle = C d_{\sigma_1}^\dag(\bfR_1) d_{\sigma_2}^\dag(\bfR_2) \ldots d_{\sigma_N}^\dag(\bfR_N) | 0 \rangle,
    \label{eq:wc_ansatz}
    \ee
where the operator 
    \be
    d_{\sigma_n}^\dag(\bfR_n) = \frac{1}{\sqrt A} \sum_\bfk e^{-i\bfk\cdot\bfR} \tilde\Phi(\bfk) c^\dag_{\bfk\sigma_n}
    \ee
creates an electron with spin $\sigma_n$ localized at the WC lattice site $\bfR_n$. The corresponding localized wave function is 
$\Phi(\bfr) =  \exp(-r^2/4\sigma^2) / \sqrt{2\pi \sigma^2}$, with Fourier transform $\tilde\Phi(\bfk)$, and the width $\sigma$ is a variational parameter. Constant $C$ is a normalization factor. 

Here and below, we present our results for the variational energies in the large-$r_s$ limit, where the exponentially small overlap of Gaussian orbitals localized at different WC lattice sites can be neglected \footnote{As shown in \cite{supp}, this is equivalent to the conventional Hartree approximation, which, for $\Omega = 0$, gives an approximation of the energy that is within 10\% of more accurate quantum Monte Carlo calculations \cite{drummond2009}.}. More detailed formulas and  derivations are given in \cite{supp}. In the large-$r_s$ limit, the energy per electron of the ordinary WC in the state \eqref{eq:wc_ansatz} is independent of the spin configuration and is found to be
	\be
    \frac{E^{\text{WC}}_\text{var}(\sigma)}{N} \approx \frac{\hbar^2}{4m^{*}\sigma^2} + \epsilon_\text{cl} + \frac{\gamma e^2\sigma^2}{2a_\text{WC}^3} \left(1 + \frac{\Omega}{4\sigma^2}\right)^{2},
    \ee
where $\epsilon_\text{cl} = \beta e^2/a_\text{WC}$ with $\beta \approx -2.107$ is the classical electrostatic energy (per electron) of a triangular lattice \cite{bonsall1977}, and $\gamma \approx 11.03$ is obtained by evaluating a triangular lattice dipolar sum \cite{supp}.  
Minimizing with respect to $\sigma$ yields the variational energy
    \be
    \begin{aligned}
    &\frac{E^\text{WC}_\text{var}}{N} = \frac{A}{r_s} + \frac{B(\omega)}{r_s^{3/2}} ~ \text{Ha}. 
    \end{aligned}
    \ee
Here $A \approx -1.106$, $B(\omega) \approx 0.40 \omega + 0.89\sqrt{1+0.20\omega^2}$, and the energy is in Hartree units $e^2/a_B$. These results coincide with those obtained in \cite{tan2025}. 

Now we consider the \textit{paired} WC, for which we use a variational state corresponding to a product of pairs:
    \be
    |\Psi_p \rangle = C_p b^\dag(\bfR_1)b^\dag(\bfR_2) \ldots b^\dag(\bfR_{N_p}) | 0\rangle,
    \label{eq:ansatz_paired}
    \ee
where $N_p=N/2$ is the number of pairs and the operator
    \be
    b^\dag(\bfR_n) = \frac{1}{A}\sum_{\bfk\bfq}e^{-i\bfq\cdot\bfR_n} \tilde\Phi(\bfq) \tilde\varphi(\bfk)\chi_{\sigma\sigma'}c^\dag_{\bfq/2+\bfk\sigma}c^\dag_{\bfq/2-\bfk\sigma'}
    \ee
    creates a pair localized at the WC lattice site $\bfR_n$. Here $\tilde\Phi(\bfq)$ and $\tilde\varphi(\bfk)$ are Fourier transforms of the center-of-mass (COM) and relative wave functions of the pair, respectively. We take the COM wave function $\Phi(\bfR)$ to be a Gaussian, just like that for the ordinary WC \eqref{eq:wc_ansatz}, with a variational width $\sigma_p$. The spin part of the wave function is $\chi_{\sigma\sigma'}$, and constant $C_p$ is the normalization factor. In the large-$r_s$ limit, where the overlap between pairs at different sites may again be neglected, the operators $b^\dag(\bfR_n)$ effectively behave as hardcore bosons.
Using the state \eqref{eq:ansatz_paired}, the energy per pair in the large-$r_s$ limit is
    \be
\begin{aligned}
	&\frac{E^\text{PWC}_\text{var}(\sigma_p)}{N_{p}} \approx \frac{\hbar^2}{4M \sigma_p^2} +\epsilon'_\text{cl}+ \frac{2\gamma e^{2}\sigma_p^2}{a_\text{WC}^{'3}}\left(1 + \frac{\Omega}{8\sigma_p^2}\right)^{2} +\epsilon_{0m}.
	\end{aligned}
    \label{eq:paired_energy_var}
    \ee
    Here $M = 2m^{*}$ is the total mass of the pair, $a'_\text{WC} = \sqrt{2} a_\text{WC}$ is the lattice constant of the paired crystal, $\epsilon'_\text{cl} = \beta (2e)^2 / a'_\text{WC}$, and $\epsilon_{0m}(r_s,\omega)$ is the lowest-energy eigenvalue of a two-electron pairing problem, which depends on the relative angular momentum $m$ of the pair. For large $r_s$, $\epsilon_{0m}$ is independent of $\sigma_p$. 
Minimizing with respect to $\sigma_p$ and converting to energy per particle gives the variational energy 
    \be
    \begin{aligned}
    \frac{E^\text{PWC}_\text{var}}{N} = \frac{A_p}{r_s} + \frac{B_p(\omega)}{r_s^{3/2}} + \frac{\epsilon_{0m}(r_s, \omega)}{2} ~ \text{Ha},
    \end{aligned}
    \ee
    where $A_p =\sqrt{2} A  \approx -1.564$, $B_p(\omega) \approx 0.14 \omega  + 0.38 \sqrt{1+0.14\omega^2}$, and the energy eigenvalue $\epsilon_{0m}$ is obtained by solving the Schrödinger equation for the relative motion of the electron pair in a unit cell. Writing the wave function for the relative coordinate in $k$-space as $\tilde\varphi(\bfk)= \phi_m(k)e^{im\theta}$, the Schrödinger equation projected onto the angular momentum sector $m$ reads
\begin{widetext}
\begin{align}
     & \left\{\frac{\hbar^{2}k^{2}}{2\mu}-\frac{K}{2}\left[ \frac{d^{2}}{dk^{2}}+\frac{1}{k}\frac{d}{d k}-\frac{1}{k^{2}}\left(m-\frac{\Omega_{r}k^{2}}{2}\right)^{2} \right] \right\}\phi_{m}\left(k\right) +\frac{K\Omega_r}{2}\phi_{m}\left(k\right) +\int^{\infty}_{0} V_{m}\left(k,k^{\prime}\right)\phi_{m}\left(k^{\prime}\right)k^{\prime}dk^{\prime}=\epsilon_{nm}\phi_{m}\left(k\right),
     \label{eq:Schrodinger_momentum}
\end{align}
\end{widetext}
where $\mu = m^{*}/2$ is the reduced mass of the pair, $\Omega_r = 2 \Omega$ (the factor of 2 arises from translating to relative and COM coordinates), and the Coulomb interaction is dressed with form factors:
    \be
V_{m}\left(k,k^\prime\right)=e^{2}\int^{2\pi}_{0}\frac{e^{-im\theta}e^{\frac{\Omega_r}{2}kk^{\prime}e^{i\theta}}e^{-\frac{\Omega_r}{4}\left(k^{2}+k^{\prime 2}\right)}}{\sqrt{k^{2}+k^{\prime 2}-2kk^{\prime}\cos{\theta}}}\frac{d\theta}{2\pi}.
    \label{eq:dressed_coulomb}
    \ee
The intracell harmonic confining potential in the Schrödinger equation \eqref{eq:Schrodinger_momentum} (which acts as a Laplacian operator in $k$-space) arises from the intercell interactions with the surrounding electrons. The corresponding stiffness $K$ has a magnitude determined by the curvature of the Coulomb potential and is given by $K = \gamma e^2/2r_s^3a_B^3$. The constant term $K\Omega_r/2$ in \eqref{eq:Schrodinger_momentum} arises from the quantum metric. The detailed derivation of the effective ``quantum dot" problem  \eqref{eq:Schrodinger_momentum} starting from the expectation value in the state \eqref{eq:ansatz_paired} is provided in \cite{supp}. 

\begin{figure}[t!]
    \centering
    \includegraphics[]{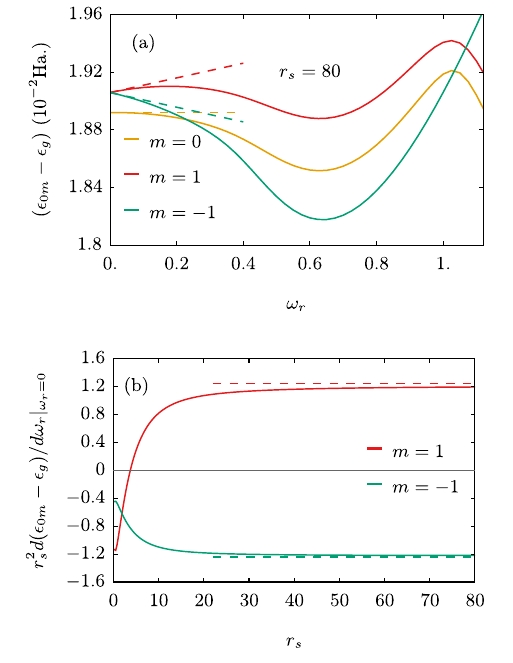}
    \caption{(a) Lowest-energy eigenvalues of the Schrödinger equation \eqref{eq:Schrodinger_momentum} as a function of Berry curvature $\omega_{r}$ for $r_s=80$ and different orbital angular momenta $m$. To highlight the difference between the curves we subtract the energy $\epsilon_g$ (see text). Dashed lines are the perturbative result \eqref{eq:de_pert}. (b) Derivative of the lowest-energy eigenvalues with respect to $\omega_r$ as a function of $r_s$ for $m=\pm 1$. Dashed lines show the asymptotic large-$r_s$ behavior \eqref{eq:de_pert}. Derivatives are multiplied by $r_s^2$ for visual clarity.}
    \label{fig:qd_energies}
\end{figure}
We have numerically diagonalized the Schrödinger equation \eqref{eq:Schrodinger_momentum}. 
The evolution of the energies for different angular momenta $m$ as a function of $\omega$ is shown in Fig.~\ref{fig:qd_energies}a for the representative value $r_s = 80$. When $\omega$ = 0, the lowest-energy state has orbital angular momentum $m = 0$, corresponding to a spin-singlet pair wave function. However, for $r_s \gtrsim 15$, the $m=0$ state only remains the ground state for sufficiently small $\omega$; increasing the Berry curvature eventually switches the ground state to $m = -1$ (anti-aligned with the direction of the Berry curvature), corresponding to a spin-triplet state. Unlike the case of singlet-triplet transitions in two-electron quantum dots in a perpendicular magnetic field, where the magnitude of the ground state orbital angular momentum increases with increasing field \cite{wagner1992,ashoori1993}, we do not find ground states with any other values of $m$. In fact, for sufficiently large $\omega$, the ground state returns to $m=0$.  
This evolution explains the shape of the $m=-1$ paired WC phase in Fig.~\ref{fig:phase_diag}. 
Interestingly, recent BCS calculations \cite{maymann2025} on the model \eqref{eq:H} also found that repulsive interactions favor pairing in which the relative angular momentum of the pair is anti-aligned with the Berry curvature.

The essential features of the quantum dot problem can already be understood from a perturbative treatment of the Berry curvature at large $r_s$. When $\Omega_{r} = 0$, the real-space Schrödinger equation for the relative motion is
    \be
    \left[-\frac{\hbar^2}{2\mu} \frac{d^2}{dr^2} + \frac{\hbar^2(m^2-1/4)}{2\mu r^2} + V_\text{QD}(r)\right] u(r) = \epsilon_{nm} u(r),
    \label{eq:rad_eq}
    \ee
where $u(r) = \sqrt r \phi(r)$ and we define the effective quantum dot potential $V_\text{QD}(r) = K r^2/2 + e^2/r$. The classical equilibrium separation $r_0$ between the electrons forming the pair is determined by minimizing $V_\text{QD}'(r_0) = 0$ and is given by $r_0 = \left(e^{2}/K\right)^{1/3} \sim r_s a_B$. 

At large $r_s$, when $r_0$ is much larger than the fluctuations in the distance between electrons in the pair, we may expand about the minimum of the potential $ V_\text{QD}(r) \approx V_\text{QD}(r_0) + \frac 12 V_\text{QD}''(r_0)(r-r_0)^2$. In this approximation, the energy spectrum is that of a one-dimensional harmonic oscillator: $\epsilon_{nm} = V_\text{QD}(r_0) + \hbar \omega_\text{QD}(n+1/2) + \hbar^2m^2/2\mu r_0^2$, where $\omega_\text{QD} = \sqrt{3K/\mu}$ \footnote{In this semiclassical limit, a better approximation to the energy is obtained by substituting $m^2-1/4 \to m^2$ \cite{brack2018}.}. The corresponding wave functions for the lowest harmonic oscillator level are $u(r)\sim \exp[-(r-r_0)^{2}/2\xi^2]$, where $\xi^2 = \hbar/\sqrt{3 K \mu} \sim r_s^{3/2}a_B^2$. The ratio $\xi/r_0 \sim r_s^{-1/4}$ is indeed small when $r_s$ is large, justifying the harmonic approximation. 

We now expand \eqref{eq:Schrodinger_momentum} to leading order in $\Omega_{r}$. The correction to the quantum dot Hamiltonian \eqref{eq:rad_eq}, expressed in real space, is
    \be
    \Delta H = -\frac{\Omega_r}{2r}\frac{dV}{dr}m+\frac{\Omega_r}{4}\nabla^2 V.
    \label{eq:dH}
    \ee
The first term arises from the Berry curvature, and the second is from the quantum metric. Similar quantum-geometric correction terms to the two-body problem have appeared in the context of exciton formation \cite{srivastava2015,zhou2015} and Cooper pairing \cite{simon2022}. First order perturbation theory using the one-dimensional harmonic oscillator wave function above 
gives the energy shift
    \be
    \Delta \epsilon_{0m} =\epsilon_{g}+ \frac {3m}{4}\frac{e^2}{r_0^5}\xi^2  \Omega_r, ~ \epsilon_{g}=\frac 34 \frac{e^2}{r_0^3}\left(1 + \frac{\xi^2}{2r_0^2}\right)\Omega_r.
    \label{eq:de_pert}
    \ee
The term $\epsilon_{g}$ is the quantum metric contribution and gives a positive energy correction to all states, independent of the angular momentum. The remaining $m$-dependent term lowers the energy of states with negative angular momentum. When the interactions are sufficiently strong, so that the spacing $\hbar^2 m^2 / 2 \mu r_0^2 \propto 1 / r_s^2$ between angular momentum states is small, the energy lowering is sufficient to drive a transition to a state of nonzero angular momentum.

\begin{figure}
    \centering
    \includegraphics[]{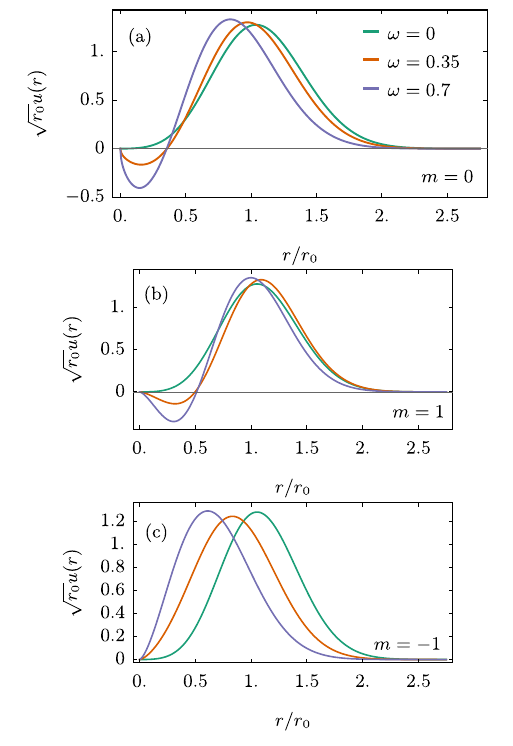}
    \caption{Berry curvature dependence of lowest-energy wave functions $u(r)$ of the effective quantum dot problem \eqref{eq:Schrodinger_momentum},  for angular momenta  (a) $m=0$, (b) $m=1$, and (c) $m=-1$. Here $r_s = 80$ and $r_0 = (e^2/K)^{1/3}$ is the classical distance between electrons in the pair; see Eq.~\eqref{eq:rad_eq}. }
    \label{fig:wave_funcs}
\end{figure}

To highlight the crucial role of strong correlations for the $m=0 \to m=-1$ transition, we show in Fig.~\ref{fig:qd_energies}b the derivatives $d(\epsilon_{0,m=\pm 1} - \epsilon_g)/d\omega_r$ at $\omega_r = 0$, obtained from the full numerical diagonalization of \eqref{eq:Schrodinger_momentum}. At small $r_s$, this derivative is more negative for $m=+1$ than for $m=-1$, and small Berry curvature therefore favors the $m=+1$ state, corresponding to a ``momentum-space orbital Zeeman effect" \cite{srivastava2015}. However, with increasing $r_s$, the $m=-1$ derivative becomes more negative (and that for $m=+1$ eventually  becomes positive), implying that the $m=-1$ state is energetically favored by small Berry curvature.

In Fig.~\ref{fig:wave_funcs}, we show the evolution  of the lowest-energy radial wave functions $u(r)$ for $m=0,\pm 1$ as the Berry curvature is increased. Several unusual features are worth noting. In all cases shown, the ground-state wave functions develop additional weight near $r=0$, corresponding to configurations in which the electrons in a pair have increased spatial overlap. While the ground-state wave function for $m=-1$ remains nodeless, the ground-state wave functions for $m=0$ and $m=+1$ develop nodes. Such behavior is forbidden for \textit{local} Schrödinger equations, whose ground states are necessarily nodeless (for each $m$), and appears here as a consequence of the nonlocal nature of the Coulomb potential in  Eq.~\eqref{eq:Schrodinger_momentum}: For nonzero $\Omega_r$, the potential energy acts as an integral operator, and the total energy can be lowered by the development of oscillations in the wave function---even in classically forbidden regions---despite the associated increase in kinetic energy. Qualitatively, this nodal structure suggests that the energetic favoring of the $m=-1$ state arises because it can take advantage of oscillations already present in the Coulomb kernel in Eq.~\eqref{eq:Schrodinger_momentum}, without incurring the additional kinetic energy cost associated with introducing nodes. A more detailed analysis of the wave functions, however, lies beyond the scope of this work. 

Finally, we comment on the expected effects of an applied magnetic field. The Zeeman effect favors the spin-triplet, $m=-1$ paired WC over the spin-singlet, $m=0$ state. Accordingly, an in-plane magnetic field, which couples primarily to the electron spin, will expand the region of the phase diagram occupied by the $m=-1$ paired WC. The orbital effect of a perpendicular magnetic field, on the other hand, is pair breaking \cite{taut2001}. A sufficiently strong out-of-plane field will drive a transition from the paired WC back to the monatomic WC, potentially proceeding via additional singlet-triplet transitions \cite{taut2001}.

The variational calculations presented here can be quantitatively improved in several ways. The states \eqref{eq:wc_ansatz} and \eqref{eq:ansatz_paired} correspond to an ``Einstein model" of the electron solid. For the monatomic WC, systematic semiclassical calculations that incorporate the full phonon spectrum \cite{bonsall1977} provide a more quantitatively reliable estimate of the energy \cite{drummond2009}. Such calculations are also possible for paired WCs \cite{taut2001} and may yield more accurate energies than those reported here. Recently developed neural-network variational Monte Carlo methods for studying electron gases with quantum geometry \cite{valenti2025} offer another route to search for paired WC phases. More detailed calculations that account for realistic features of actual devices, such as gate screening \cite{valenti2025gate} and sample disorder \cite{xiang2025,joy2025disorder}, will be important for assessing the relevance of our results to real materials. It would also be interesting to explore connections between our findings and recently discovered Wigner molecular crystals in moiré materials \cite{li2024}. 

It is tempting to speculate that the local spin-triplet pairing tendency identified here may survive the partial or complete melting of the electron crystal. Recent calculations have hinted at the possibility of intermediate metallic density-wave states between the (monatomic) WC and the homogeneous liquid, driven by a self-doping instability of the crystal \cite{kim2024}. In the present context, the analogous state---in which the paired crystal spontaneously generates a finite concentration of (paired) ground-state defects---would correspond to a charged, bosonic supersolid, in which superfluidity coexists with translational symmetry breaking. On the other hand, a complete loss of translational order could lead to a spin-triplet superconducting state driven by strong correlations and quantum geometry, a purely electronic mechanism of superconductivity that is complementary to the more conventional Kohn--Luttinger mechanism \cite{kohn1965,chubukov1993,geier2025,shavit2025,jahin2025,chou2025,yang2025,qin2025,parra2025,maymann2025}. Exploring possible connections between this scenario and recent observations of chiral superconductivity in rhombohedral graphene \cite{han2025} is an intriguing direction for future study. 

We thank Trithep Devakul, Elio König, Julian May-Mann, Daniel Parker, Brian Skinner, and Maxim Vavilov for valuable discussions and feedback.
This research was supported by the National Science Foundation (NSF) through the University of Wisconsin Materials Research Science and Engineering Center Grant No. DMR-2309000 (I. E. and D. Z.) and NSF Grant No. DMR-2452658 (A. L.). Support for this research was also provided by the Office of the Vice Chancellor for Research and Graduate Education at the University of Wisconsin–Madison with funding from the Wisconsin Alumni Research Foundation, as well as from the University of Wisconsin – Madison (I. E.) and H. I. Romnes Faculty Fellowship (A. L.).

\bibliography{paired_crystal_berry}

\clearpage

\widetext

\begin{center}
{\large \bf Supplemental Material for\\
\textit{``Spin-triplet paired Wigner crystal stabilized by quantum geometry''}}
\end{center}

\section{Variational states and energies: second-quantized approach}

In this supplemental section we provide further details on the variational states used for the monatomic and paired Wigner crystals (WCs), including certain useful operator identities for the $d^\dag$ and $b^\dag$ operators introduced in the main text, as well as more detailed calculations of the variational energies. 

\subsection{Variational states and operator identities}

For reference, we record here the Hamiltonian given in Eq.~(1) of the main text:
    \be
    H = \sum_{\bfk\sigma} \frac{\hbar^2k^2}{2m^*}c^\dag_{\bfk\sigma}c_{\bfk\sigma} +\frac{1}{2A}\sum_{\bfq \neq 0} v_\bfq \sum_{\substack{\bfk\bfk' \\ \sigma \sigma'}}
    F_{\bfk+\bfq,\bfk}F_{\bfk'-\bfq,\bfk'}c^\dag_{\bfk+\bfq\sigma}c^\dag_{\bfk'-\bfq\sigma'}c_{\bfk'\sigma'}c_{\bfk\sigma},
    \label{eq:H}
    \ee
where $v_\bfq = 2\pi e^2/q$ is the Fourier transform of the Coulomb interaction, $A$ is the area of the system, and the form factors are
    \be
    F_{\bfk, \bfk'}= e^{-i\frac{\Omega}{2}\bfk \times \bfk'}e^{-\frac{\Omega}{4}|\bfk - \bfk'|^2}.
    \label{eq:form_factor}
    \ee

For the ordinary, monatomic $N$-electron WC we use the variational state 
	\be
	| \Psi \rangle = C d^\dag_{\sigma_1}(\bfR_1) \ldots d^\dag_{\sigma_N}(\bfR_N) | 0 \rangle, \quad d^\dag_\sigma(\bfR_n) = \frac{1}{\sqrt A} \sum_\bfk e^{-i\bfk \cdot \bfR_n} \tilde \Phi(\bfk) c^\dag_{\bfk \sigma},
	\label{eq:wc_anstaz_rspace}
	\ee
where $\tilde\Phi(\bfk)$ is the Fourier transform of the localized orbital $\Phi(\bfr) = \exp(-r^2/4\sigma^2)/\sqrt{2\pi\sigma^2}$. This type of state was used to describe monatomic WCs in Refs.~\cite{vandijk1991,moulopoulos1993,bonev2008}. 
Operators $d^\dag_\sigma(\bfR_n)$ have the interpretation of creating an electron at WC lattice site $\bfR_n$ with spin $\sigma_n$. The anticommutation relations are 
	\be
	\{d_{\sigma}(\bfR),d^\dag_{\sigma'}(\bfR')\} = \delta_{\sigma\sigma'} S(\bfR - \bfR'), \quad \{d_{\sigma}(\bfR),d_{\sigma'}(\bfR')\} = \{d^\dag_{\sigma}(\bfR),d^\dag_{\sigma'}(\bfR')\} = 0.
	\ee
Function $S$ measures overlaps between localized orbitals:
    \be
    S(\bfR - \bfR') = \int \dd^2\bfr ~ \Phi^*(\bfr - \bfR) \Phi(\bfr - \bfR') = e^{-(\bfR - \bfR')^2/8\sigma^2}.
    \ee
Throughout we will neglect this exponentially small wave function overlaps between different sites and make the approximation $S(\bfR - \bfR') \approx \delta_{\bfR \bfR'}$, so that the $d^\dag$ operators obey standard fermion anticommutation relations. In this approximation, the normalization constant $C \approx 1$. Higher-order terms in a systematic expansion in powers of $S$ can be found in Refs.~\cite{moulopoulos1993, vandijk1991,bonev2008}. Below we will also make use of the anticommutation relation
	\be
	\{ c_{\bfk\sigma}, d^\dag_{\sigma'}(\bfR_n) \} = \frac{1}{\sqrt A} \delta_{\sigma\sigma'} e^{-i\bfk \cdot \bfR_n} \tilde \Phi(\bfk).
	\ee
For the paired WC we use the state 
	\be
	|\Psi_p \rangle = C_p b^\dag(\bfR_1) b^\dag(\bfR_2) \ldots b^\dag(\bfR_{N_p})| 0 \rangle, 
    \quad b^\dag(\bfR_n) = \frac 1A \sum_{\bfk\bfq \sigma\sigma'} e^{-i\bfq \cdot \bfR_n} \tilde \Phi(\bfq) \tilde \varphi(\bfk)  \chi_{\sigma \sigma'} c^\dag_{\bfq/2+\bfk, \sigma} c^\dag_{\bfq/2-\bfk, \sigma'},
    \label{eq:paired_ansatz}
	\ee
where $N_p = N/2$ is the number of pairs. 
The spatial part of the pair wave function localized at site $\bfR_n$ is
    \be
    \psi_n(\bfr, \bfr') = \Phi\left(\frac{\bfr + \bfr'}{2} - \bfR_n\right) \varphi(\bfr - \bfr'),
    \ee
where $\bfr$ and $\bfr'$ are the coordinates of the two electrons comprising the pair. Functions $\Phi$ and $\varphi$ are the center-of-mass (COM) and relative wave functions, respectively. The spin part of the wave function is $\chi_{\sigma\sigma'}$. For singlet pairing $\chi_{\sigma\sigma'} = (\delta_{\sigma\uparrow}\delta_{\sigma'\downarrow}-\delta_{\sigma\downarrow}\delta_{\sigma'\uparrow})/\sqrt 2$, while for triplet pairing the three possible spin states are $\chi^{(+1)}_{\sigma\sigma'} = \delta_{\sigma\uparrow}\delta_{\sigma'\uparrow}$, $\chi^{(0)}_{\sigma\sigma'} = (\delta_{\sigma\uparrow}\delta_{\sigma'\downarrow}+\delta_{\sigma\downarrow}\delta_{\sigma'\uparrow})/\sqrt 2$, and $\chi^{(-1)}_{\sigma\sigma'} = \delta_{\sigma\downarrow}\delta_{\sigma'\downarrow}$. The state \eqref{eq:paired_ansatz} was used earlier in Ref.~\cite{moulopoulos1993} to describe spin-singlet paired WCs in the three-dimensional electron gas. 

Below will make use of the following identities involving the $b^\dag$ operators:
    \begin{align}
    [c^\dag_{\bfk\sigma}c_{\bfk'\sigma'}, b^\dag(\bfR_n) ] &= \frac 2A \sum_{\bfk''\sigma''}e^{-i(\bfk' + \bfk'')\cdot \bfR_n} \tilde\Phi(\bfk' + \bfk'') \tilde\varphi\left(\frac{\bfk' - \bfk''}{2}\right) \chi_{\sigma' \sigma''} c^\dag_{\bfk\sigma} c^\dag_{\bfk''\sigma''}, \\
	[b(\bfR_n), [c^\dag_{\bfk\sigma}c_{\bfk'\sigma'}, b^\dag(\bfR_n) ] ] |0 \rangle &= \frac{4}{A^2} \sum_{\bfk''\sigma''} e^{i(\bfk-\bfk')\cdot \bfR_n} \tilde \Phi^*(\bfk + \bfk'')\tilde \Phi(\bfk' + \bfk'') \tilde\varphi^*\left(\frac{\bfk - \bfk''}{2}\right) \tilde\varphi\left(\frac{\bfk' - \bfk''}{2}\right)  \chi_{\sigma \sigma''}\chi_{\sigma' \sigma''} | 0 \rangle,
	\label{eq:nested_comm}
    \end{align} 
as well as the approximation relations 
    \begin{align}
    [b(\bfR_n),b^\dag(\bfR_m)] &\approx 0 \quad (m\neq n), 
    \label{eq:b_comm}\\
    b(\bfR_n)b^\dag(\bfR_n)|0 \rangle &\approx 2 |0\rangle,
    \label{eq:b_norm}
    \end{align}
which are valid in the limit of vanishing intercell overlaps. In this approximation, the constant $C_p \approx 2^{-N_p/2}$. 

In the following subsections, we compute expectation values of the Hamiltonian \eqref{eq:H} using the states \eqref{eq:wc_anstaz_rspace} and  \eqref{eq:paired_ansatz}. As described above, we neglect the exponentially small overlaps of wave functions in different unit cells, which gives corrections of the order $\sim \exp(-\alpha a_\text{WC}^2/\sigma^2)$, where $\sigma$ is the width of the localized orbital, $a_\text{WC}$ is the WC lattice constant ,  and $\alpha = \mathcal O(1)$. In the next section of this supplement, we demonstrate that this approximation is equivalent to the ``Hartree approximation" in a first-quantized, wave function language. At various stages of the calculations below, we will make use of the usual passage from discrete to continuous momenta $A^{-1}\sum_\bfk \to \int \dd^2\bfk /(2\pi)^2$.

\subsection{Variational energy of the monatomic WC}

We start with the monatomic WC and adopt the shorthand notation $d^\dag(\bfR_n) \to d^\dag_n$. In the approximation of neglected intercell overlaps, the energy of the monatomic WC is independent of the spin configuration and we will therefore suppress the spin labels in what follows.
The expectation value of the kinetic energy is	$ \langle \Psi | T | \Psi \rangle = \sum_\bfk \epsilon_\bfk \langle \Psi |  c^\dag_\bfk c_\bfk | \Psi \rangle $. Expanding the expectation value
	\begin{align}
	\langle \Psi |  c^\dag_\bfk c_\bfk | \Psi \rangle &= \langle 0 | d_N \ldots d_1 c^\dag_\bfk c_\bfk d^\dag_1 \ldots d^\dag_N | 0\rangle \\
	&= \sum_i (-1)^{i+1} \{ c_\bfk, d^\dag_i \} \langle 0 | \prod_m d_m c^\dag_\bfk \prod_{n\neq i} d^\dag_n | 0 \rangle \\
	&= \sum_{ij} (-1)^{i+j} \{ c_\bfk, d^\dag_i \} \{d_j, c^\dag_\bfk \}  \langle 0 | \prod_{m\neq j} d_m  \prod_{n\neq i} d^\dag_n | 0 \rangle \\
	&\approx \sum_i \{ c_\bfk, d^\dag_i \} \{d_i,  c^\dag_\bfk \} \\ 
	&= \frac{N}{A} \tilde \Phi^*(\bfk) \tilde \Phi(\bfk) 
	\end{align}
The second to last line is obtained using the approximation of neglected overlaps \eqref{eq:b_comm}. We thus find
    \be
    \langle \Psi | T | \Psi \rangle = N \int \frac{\dd^2\bfk}{(2\pi)^2} ~  \epsilon_\bfk \tilde \Phi^*(\bfk) \tilde\Phi(\bfk) = N\frac{\hbar^2}{4m^*\sigma^2}.
    \label{eq:t_wc}
    \ee
The potential energy is obtained similarly. We have
	\be
	 \langle \Psi | V | \Psi \rangle  = \frac{1}{2A}  \sum_{\bfq \neq 0} v_\bfq \sum_{\bfk \bfk'} F_{\bfk+\bfq, \bfk} F_{\bfk'-\bfq, \bfk'}  \langle \Psi | c^\dag_{\bfk+\bfq}c^\dag_{\bfk'-\bfq} c_{\bfk'} c_{\bfk} | \Psi \rangle.
	\ee
Neglecting intercell overlaps, one obtains
	\be
	 \langle \Psi | c^\dag_{\bfk+\bfq}c^\dag_{\bfk'-\bfq} c_{\bfk'} c_{\bfk} | \Psi \rangle = \frac{1}{A^2}\sum_{i \neq j} e^{i\bfq \cdot (\bfR_i - \bfR_j)} \tilde \Phi^*(\bfk + \bfq) \tilde \Phi^*(\bfk' - \bfq) \tilde \Phi(\bfk') \tilde \Phi(\bfk) .
	 \ee
The manipulations required to arrive at this result are similar to those made in calculating the kinetic energy,  with additional oscillatory terms of the form 
	\be
	\int \frac{\dd^2\bfk}{(2\pi)^2} ~ F_{\bfk+\bfq, \bfk}\tilde \Phi^*(\bfk+\bfq) \tilde \Phi(\bfk) e^{-i\bfk \cdot (\bfR - \bfR')} \sim e^{-(\bfR - \bfR')^2/8\sigma^2} ,
	\ee
which are again exponentially small in proportion to the overlap and are thus dropped. We obtain
	 \be
	 \langle \Psi | V | \Psi \rangle = \frac{1}{2A} \sum_{i \neq j } \sum_{\bfq \neq 0 }   v_\bfq e^{i\bfq \cdot (\bfR_i - \bfR_j)}  \left| \int \frac{\dd^2\bfk}{(2\pi)^2} ~ F_{\bfk - \bfq,\bfk} \tilde \Phi^*(\bfk-\bfq) \tilde \Phi(\bfk) \right|^2.
	\ee
Defining the electron density $n(\bfq)=  \langle \Psi | \sum_\bfk F_{\bfk-\bfq,\bfk} c^\dag_{\bfk-\bfq} c_\bfk | \Psi \rangle$, we have, in the approximation of vanishing overlaps,
	\be
	n(\bfq) =  \sum_j e^{-i \bfq \cdot \bfR_j} n_0(\bfq), \quad n_0(\bfq) \equiv \int \frac{\dd^2\bfk}{(2\pi)^2}~  F_{\bfk - \bfq,\bfk} \tilde \Phi^*(\bfk-\bfq) \tilde \Phi(\bfk) =  e^{-\frac{\sigma^2}{2}\left(1 + \frac{\Omega}{4\sigma^2}\right)^2 q^2}.
	\label{eq:bc_density}
	\ee
 The potential energy term can then be written
	\be
	\langle \Psi | V | \Psi \rangle = \frac{N}{2A} \sum_{\bfR_n\neq 0} \sum_{\bfq \neq 0} v_\bfq  | n_0(\bfq) |^2 e^{i\bfq \cdot \bfR_n}. 
    \ee
We now convert the sum on $\bfq$ to an integral, keeping in mind that we must also then add in the interaction with the neutralizing background. Evaluating the integral (and keeping the background term implicit) gives
    \be
    \langle \Psi | V | \Psi \rangle = \frac{N}{2}e^2 \sum_{\bfR_n \neq 0} \frac{\sqrt \pi}{2 \tilde\sigma} e^{-R_n^2/8\tilde\sigma^2}I_0\left(\frac{R_n^2}{8\tilde\sigma^2}\right),
    \label{eq:V_WC_I0}
    \ee
where $\tilde \sigma^2 = \sigma^2 (1+\Omega/4\sigma^2)^2$ and $I_n(x)$ is the modified Bessel function. 

For large $r_s$, the the localized orbitals will be sharply peaked around the lattice sites, $\sigma \ll a_\text{WC}$, and we may expand
	\be
	\frac{\sqrt \pi}{2\tilde \sigma} e^{-R_n^2/8\tilde \sigma^2} I_0\left(\frac{R_n^2}{8\tilde \sigma^2}\right) = \frac{1}{R_n} + \frac{\tilde\sigma^2}{R_n^3} + \ldots.
	\ee
To the lowest order, we thus arrive at the result
	\be
	\langle  \Psi | V | \Psi \rangle = \frac N2 \sum_{\bfR_n \neq 0} \frac{e^2\tilde\sigma^2}{R_n^3}+ E_\text{cl}, \quad E_\text{cl} = \frac N2 \sum_{\bfR_n \neq 0} \frac{e^2}{R_n}.
    \label{eq:V_WC}
	\ee
Here $E_\text{cl}$ is the classical Coulomb energy of the crystal, which implicitly includes the interaction with the neutralizing background \cite{bonsall1977}. The dipolar sum may be evaluated numerically: $\sum_{\bfR_n \neq 0} 1/R_n^3 = \gamma / a_\text{WC}^3$, where $\gamma \approx 11.03$. 

Combining the potential energy with the kinetic energy \eqref{eq:t_wc} we arrive at Eq.~(5) of the main text for the energy per electron of the monatomic WC. We note that this result for the energy of the WC in a band with quantum geometry has also been obtained in \cite{tan2025}.

\subsection{Variational energy of the paired WC}

We now turn to the paired WC and use the same shorthand as above $b^\dag(\bfR_n) \to b^\dag_n$. The expectation value of the kinetic energy is
	\begin{align}
	\langle \Psi_p | T | \Psi_p \rangle &= \frac {1}{2^{N_{p}}}\sum_{\bfk\sigma} \epsilon_\bfk \langle 0 | b_{N_p} \ldots  b_1 (c^\dag_{\bfk\sigma} c_{\bfk\sigma} ) b^\dag_1 \ldots b^\dag_{N_p} | 0\rangle \\
	&=  \frac {1}{2^{N_{p}}}\sum_{\bfk\sigma} \epsilon_\bfk \langle 0 | b_{N_p} \ldots  b_2 (b_{1}  b^\dag_{1}) (c^\dag_{\bfk\sigma} c_{\bfk\sigma} ) b^\dag_2 \ldots b^\dag_{N_p} | 0\rangle  + \frac{1}{2^{N_{p}}} \sum_{\bfk\sigma} \epsilon_\bfk \langle 0 | b_{N_p} \ldots  b_{2} ( b_1 [c^\dag_{\bfk\sigma} c_{\bfk\sigma}, b^\dag_1] ) b^\dag_2 \ldots b^\dag_{N_p} | 0\rangle 
	\end{align}
Next, we write $b_1[c^\dag_{\bfk\sigma} c_{\bfk\sigma}, b^\dag_1]  = [b_1, [c^\dag_{\bfk\sigma} c_{\bfk\sigma}, b^\dag_1] ] +   [c^\dag_{\bfk\sigma} c_{\bfk\sigma}, b^\dag_1] b_1$ and use the (approximate) bosonic commutation relations \eqref{eq:b_comm} to move operator $b_1$ in the second term all the way to the right to annihilate the vacuum. Utilizing the nested commutator \eqref{eq:nested_comm}, we thus obtain
\begin{align}
	\langle \Psi_p | T | \Psi_p \rangle &= \frac{1}{2^{N_p}} \sum_{\bfk\sigma} \epsilon_\bfk \langle 0 | b_{N_p}\ldots  b_2 (b_1  b^\dag_1) (c^\dag_{\bfk\sigma} c_{\bfk\sigma} ) b^\dag_2 \ldots b^\dag_{N_p} | 0\rangle  \nonumber \\
	&\quad +  \frac{4}{A^2} \sum_{\bfk\bfk'} \epsilon_\bfk |\Phi(\bfk + \bfk')|^2 \left |\varphi\left(\frac{\bfk - \bfk'}{2}\right) \right|^2 
	\frac{1}{2^{N_p}}\underbrace{ \langle 0 |b_{N_p} \ldots  b_2 b^\dag_2 \ldots b^\dag_{N_p} | 0\rangle}_{\approx 2^{N_p-1}} .
	\end{align}
Continuing to commute the operator $c^\dag_{\bfk\sigma} c_{\bfk\sigma}$ to the right gives $N_p$ similar terms, resulting in
	\begin{align}
	\langle \Psi_p | T | \Psi_p \rangle  &=  2 N_p \int \frac{\dd^2\bfk}{(2\pi)^2} ~ \frac{\dd^2\bfk'}{(2\pi)^2} \epsilon_{\bfk} |\Phi(\bfk + \bfk')|^2 \left |\varphi\left(\frac{\bfk - \bfk'}{2}\right) \right|^2   \\
    &= N_p  \int \frac{\dd^2\bfq}{(2\pi)^2} ~  \frac{\hbar^2\bfq^2}{2M} |\Phi(\bfq)|^2 + N_p \int \frac{\dd^2\bfk}{(2\pi)^2} ~  \frac{\hbar^2\bfk^2}{2\mu} |\varphi(\bfk)|^2,
    \label{eq:T_PWC}
	\end{align}
where $M=2m^*$ and $\mu = m^*/2$. The kinetic energy thus factorizes into a sum of COM and relative contributions. Using the Gaussian form of the COM wave function, we have explicitly $T_\text{COM} = N_p \hbar^2 /4 M\sigma_p^2$.

The calculation of the potential energy is more involved, but the basic manipulations are very similar to those made in calculating the kinetic energy. Here we just highlight the main steps. The expectation value is
	\be
	\langle \Psi_p | V | \Psi_p \rangle = \frac{1}{2A} \sum_{\bfq \neq 0} v_\bfq \sum_{\bfk\bfk'} F_{\bfk + \bfq,\bfk}F_{\bfk' - \bfq,\bfk'}  \frac{1}{2^{N_p}}\langle 0 | b_{N_p} \ldots  b_1 \mathcal O^{\bfq}_{\bfk\bfk'} b^\dag_{1} \ldots b^\dag_{N_p} | 0\rangle,
	\ee
where $\mathcal O^{\bfq}_{\bfk \bfk'} \equiv c^\dag_{\bfk'-\bfq\sigma'}c^\dag_{\bfk+\bfq\sigma}c_{\bfk\sigma}c_{\bfk'\sigma'}$. As with the kinetic energy, we express this in terms of nested commutators:
	\begin{align}
     \langle 0 | b_{N_p} \ldots  b_1 \mathcal O^{\bfq}_{\bfk\bfk'} b^\dag_1 \ldots  b^\dag_{N_p}| 0\rangle &=\sum_{n=1}^{N_p} \langle 0 | \left( \prod_{i<n} b_i b_i^\dag\right) \left( \prod_{i>n} b_i\right) \mathcal C^{\bfq}_{\bfk\bfk',n}\left( \prod_{i>n} b^\dag_i\right) | 0 \rangle,   \quad \mathcal C^{\bfq}_{\bfk\bfk',n} \equiv [b_n, \mathcal [\mathcal O^{\bfq}_{\bfk\bfk'}, b^\dag_n]].
	 \label{eq:nested_V}
	\end{align}
This result is obtained using the approximate relation \eqref{eq:b_comm}. The nested commutator $\mathcal C$ has the schematic structure
	\be
	\mathcal C^{\bfq}_{\bfk\bfk',n}  = A_n + \sum_{\alpha\beta}B_n^{\alpha\beta}c^\dag_\alpha c_\beta + \sum_{\alpha\beta\gamma\delta}D_n^{\alpha\beta\gamma\delta} c^\dag_\alpha c^\dag_\beta c_\gamma c_\delta.
	 \label{eq:commutator_decomp}
	\ee
Here $A_n$, $B_n$, and $D_n$ are $c$-numbers and their dependence on $\bfk$, $\bfk'$, and $\bfq$ has been left implicit. One can show that the four-fermion $D_n$ term gives contributions that vanish in the approximation of no intercell overlap. Direct calculation shows that the constant term $A_n$ is given by
	\be
	A_n = \frac{4}{A^2} |\tilde\Phi(\bfk + \bfk')|^2  \tilde\varphi^*\left(\frac{\bfk - \bfk'}{2} + \bfq \right) \tilde\varphi\left(\frac{\bfk - \bfk'}{2}\right)
	\ee
and is independent of the site $\bfR_n$. There will be one such term for each pair, yielding the contribution
	\be
	\langle \Psi_p | V | \Psi_p \rangle \ni V_\text{intra}  = N_p \int \frac{\dd^2\bfq}{(2\pi)^2} ~ v_\bfq e^{-\Omega q^2/2} \int \frac{\dd^2\bfk}{(2\pi)^2}~ \tilde\varphi^*(\bfk + \bfq) \tilde\varphi(\bfk) e^{-i\Omega \bfq \times \bfk}.
    \label{eq:Vintra_PWC}
	\ee
These terms corresponds to the \textit{intracell} Coulomb interaction energy of the pair.

When the bilinear term in Eq.~\eqref{eq:commutator_decomp} is inserted into \eqref{eq:nested_V}, the result can, in the approximation of vanishing intercell overlaps, again be written as a sum of expectation values of the nested commutators
    \be
    C^{\bfq}_{\bfk\bfk',nm} \equiv [b_m, [\sum_{\alpha\beta} B_n^{\alpha\beta} c^\dag_\alpha c_\beta,b_m^\dag]].
    \ee
The contribution to the expectation value from the bilinear terms is then
    \begin{align}
    \langle 0 | b_{N_p} \ldots  b_1 \mathcal O^{\bfq}_{\bfk\bfk'} b^\dag_1 \ldots  b^\dag_{N_p}| 0\rangle_\text{bilinear} &= \sum_{n>m} \langle 0 | \left( \prod_{i<n} b_i b_i^\dag\right) \left( \prod_{n<i<m} b_i b_i^\dag\right) \left( \prod_{i>m} b_i\right) \mathcal C^{\bfq}_{\bfk\bfk',nm}\left( \prod_{i>m} b^\dag_i\right) | 0 \rangle \\
    &\approx 2^{N_p-2} \sum_{n > m} \langle 0 | C^{\bfq}_{\bfk\bfk',nm} | 0 \rangle ,
    \end{align}
the last line being valid in the no-overlap approximation. 
Furthermore, direct calculation shows that only the following two types of bilinear terms survive in the no-overlap approximation:
	\be
	\begin{aligned}
	 \sum_{\alpha\beta}B_n^{\alpha\beta}c^\dag_\alpha c_\beta &\to \frac{4}{A^2}\sum_{\bfk''} e^{i\bfq \cdot \bfR_n} \tilde\Phi^*(\bfk''+\bfq) \tilde\Phi(\bfk'') \tilde\varphi^*\left(\frac{\bfk'' - 2\bfk - \bfq}{2}\right) \tilde\varphi\left(\frac{\bfk'' - 2\bfk}{2}\right) c^\dag_{\bfk'-\bfq \sigma'} c_{\bfk'\sigma'} \\
	 &\qquad +\frac{4}{A^2} \sum_{\bfk''} e^{-i\bfq \cdot \bfR_n} \tilde\Phi^*(\bfk'' - \bfq) \tilde\Phi(\bfk'') \tilde\varphi^*\left(\frac{\bfk'' - 2\bfk' + \bfq}{2}\right) \tilde\varphi\left(\frac{\bfk'' - 2\bfk'}{2}\right) c^\dag_{\bfk+\bfq \sigma} c_{\bfk\sigma}.
	 \end{aligned}
    \label{eq:bilinears}
 	\ee
Here spinor sums $\sum_{\sigma_1\sigma_2}\chi^2_{\sigma_1\sigma_2} =1$ have been evaluated.
	
Putting everything together, we arrive, after some algebraic manipulation, at the following contribution from the bilinear terms:
	\be
	\langle \Psi_p | V | \Psi_p \rangle \ni  V^0_\text{inter} = \frac{N_p}{2A}\sum_{\bfR_n \neq 0}  \sum_{\bfq\neq 0} ~ 4v_\bfq |\rho_0(\bfq)|^2    |\varrho_0(\bfq/2)|^2 e^{i\bfq\cdot \bfR_n}. 
    \label{eq:v_inter}
	\ee
We have defined, respectively, the COM and relative densities as
    \be
    \rho_0(\bfq) = \int \frac{\dd^2\bfk}{(2\pi)^2} ~ F^{(\Omega/2)}_{\bfk-\bfq, \bfk} \tilde \Phi^*(\bfk - \bfq) \tilde \Phi(\bfk), \quad \varrho_0(\bfq) = \int \frac{\dd^2\bfk}{(2\pi)^2} ~ F^{(2\Omega)}_{\bfk-\bfq, \bfk} \tilde \varphi^*(\bfk - \bfq) \tilde \varphi(\bfk),
    \ee
where the superscipt on the form factor $F$ indicates the replacements $\Omega \to \Omega/2$ and $\Omega \to 2\Omega$.

The product of densities appearing in the sum \eqref{eq:v_inter} is simply related to the Fourier component of the \textit{total} density, $n(\bfq) = \langle \Psi_p | \sum_{\bfk \sigma} F_{\bfk-\bfq,\bfk}c^\dag_{\bfk-\bfq\sigma}c_{\bfk\sigma} | \Psi_p \rangle$: in the limit of vanishing intercell overlaps, we have
    \be
    n(\bfq) = \sum_n e^{-i\bfq \cdot \bfR_n} n_0(\bfq), \quad n_0(\bfq) = 2\rho_0(\bfq) \varrho_0(\bfq/2).
    \ee 
The intercell potential energy term \eqref{eq:v_inter} is thus
	\be
	V^0_\text{inter} =  \frac{N_p}{2A}\sum_{\bfR_n \neq 0}  \sum_{\bfq \neq 0} v_\bfq |n_0(\bfq)|^2   e^{i\bfq\cdot \bfR_n}.
    \label{eq:v_inter2}
	\ee

In the final step, we make a small-$\bfq$ expansion of $\varrho_0(\bfq/2)$, which amounts to a multipole expansion of the charge distribution of the pair. After some algebra, we obtain
    \be
    \varrho_0(\bfq/2) = 1 - \frac{q^2}{16}\int \frac{\dd^2\bfk}{(2\pi)^2} ~ \varphi^*(\bfk)  \left[ \left(i\nabla_\bfk + 2 \mathcal A_\bfk\right)^2  + 2\Omega  \right] \varphi(\bfk) + \ldots
    \ee
Plugging this into \eqref{eq:v_inter}, we find
    \be
    V^0_\text{inter} = \frac{N_p}{2A}\sum_{\bfR_n \neq 0} \sum_{\bfq \neq 0} ~ v^{(2e)}_\bfq |\rho_0(\bfq)|^2 e^{i\bfq\cdot \bfR_n} + N_p  \int \frac{\dd^2\bfk}{(2\pi)^2} ~ \varphi^*(\bfk)  \left[ \frac K2\left(i\nabla_\bfk + 2 \mathcal A_\bfk\right)^2  + K\Omega \right] \varphi(\bfk)
    \label{eq:v_inter_expand}
    \ee
Here $v^{(2e)} \equiv 2\pi (2e)^2/q$ and the Berry connection is $\mathcal A_\bfk = {\boldsymbol\Omega} \times \bfk /2$.
In the first term, the internal structure of the pair drops out and we obtain the intercell Coulomb interaction between composite objects (the pairs) of charge $2e$; the result is the same as that obtained for the ordinary WC \eqref{eq:V_WC} with the substitutions $N \to N_p$, $a_\text{WC} \to a'_\text{WC} = \sqrt 2 a_\text{WC}$, $e \to 2e$, and $\Omega \to \Omega/2$. 
The second term is an \textit{intracell} harmonic confining potential that arises as a result of \textit{intercell} interactions between the pair and the surrounding electrons. The stiffness $K$ of the harmonic potential is
    \be
    K = -\sum_{\bfR_n \neq 0} \frac{e^2}{2R^3}\int_0^\infty \dd x ~ x^2 e^{-\frac{\sigma_p^2}{R_n^2}\left(1 + \frac{\Omega}{8\sigma_p^2}\right)^2 x^2} J_0(x) \approx \sum_{\bfR_n \neq 0} \frac{e^2}{2R_n^3}=\frac{\gamma e^{2}}{2a^{\prime 3}_{\text{WC}}},
    \ee
where the last approximation is for the localized limit $\sigma_p \ll a_\text{WC}$. 

The COM contribution to the kinetic energy in Eq.~\eqref{eq:T_PWC}, together with the intercell interaction energy in Eq.~\eqref{eq:v_inter_expand}, gives the first three terms of the paired WC energy in Eq.~(9) of the main text. Adding the relative kinetic energy in Eq.~\eqref{eq:T_PWC} to the intracell Coulomb interaction in Eq.~\eqref{eq:Vintra_PWC} and the harmonic confining potential in Eq.~\eqref{eq:v_inter_expand} leads to the Schrödinger equation for the internal pair energy $\epsilon_{0m}$ and pair wave function $\varphi(\bfk)$ in Eq.~(11) of the main text.

\section{Variational states and energies: wave function approach}

In this section of the supplemental material, we demonstrate that a first-quantized approach based on Hartree-type wave functions reproduces the results of the preceding section obtained using second-quantization. The appeal of the first-quantized approach lies in the relative simplicity of the variational wave functions and in the calculation of the variational energies. 

We start with the first-quantized representation of the Hamiltonian:
\be
H=\sum_{i}\frac{\hbar^{2}\hat{\bfk}_{i}^{2}}{2m^{*}}+\frac{1}{2}\sum_{i\neq j}\hat{V}_{ij},
\ee
where the Coulomb operator has matrix elements
\be
\Bra{\bfk_{i}\bfk_{j}}\hat{V}_{ij}\Ket{\bfk_{i}^{\prime}\bfk_{j}^{\prime}}=\delta\left(\bfk_{i}+\bfk_{j}-\bfk_{i}^{\prime}-\bfk_{j}^{\prime}\right)v\left(\bfk_{i}-\bfk_{i}^{\prime}\right)F_{\bfk_{i},\bfk_{i}^{\prime}}F_{\bfk_{j},\bfk_{j}^{\prime}}
\ee
and the form factors are given in Eq.~\eqref{eq:form_factor}. 
The real space matrix elements of the interaction are given by 
\begin{align}
\Bra{\bfr_{1},\bfr_{2}}\hat{V}\Ket{\bfr_{1}^{\prime},\bfr_{2}^{\prime}}&=\int \frac{\dd^2\bfq}{(2\pi)^2}v_\bfq e^{-\frac{\Omega}{2}q^{2}}e^{i\bfq\cdot\left(\bfr_{1}^{\prime}-\bfr_{2}^{\prime}\right)}\delta\left(\bfr_{1}+\frac{\left(\bfq\times\boldsymbol{\Omega}\right)}{2}-\bfr_{1}^{\prime}\right)\delta\left(\bfr_{2}-\frac{\left(\bfq\times\boldsymbol{\Omega}\right)}{2}-\bfr_{2}^{\prime}\right)\label{eq:Vrrp}\\
 &= \frac{2e^2}{\pi\Omega\left|\bfr-\bfr^{\prime}\right|}e^{-(\bfr - \bfr')^2/2\Omega}e^{-i\bfr \times \bfr'/\Omega}\delta\left(\bfR-\bfR^{\prime}\right),
\end{align}
and we introduced the COM and relative coordinates $\bfr_{1,2}=\bfR\pm\bfr/2$ and $\bfr'_{1,2} = \bfR' \pm \bfr'/2$.

For the monatomic WC, we consider an $N$-electron variational wave function in the Hartree approximation:
    \be
    \Psi(\bfr_1, \bfr_2, \ldots, \bfr_N) = \prod_{n=1}^N \Phi(\bfr_n - \bfR_n), 
    \label{eq:wc_wave_func}
    \ee
where $\Phi(\bfr) = e^{-r^2/4\sigma^2}/\sqrt{2\pi \sigma^2}$. Here each electron $n$ has been identified with a WC lattice site $\bfR_n$. We are making the approximation of distinguishable electrons, which is good up to exponentially small corrections due to exchange processes in the WC.

For the paired WC, we consider the wave function
    \be
    \Psi_p(\bfr_1, \bfr_2, \ldots, \bfr_N) = \prod_{n=1}^{N_p} \Phi\left(\frac{\bfr_{na}+\bfr_{nb}}{2} - \bfR_n\right)\varphi(\bfr_{na}-\bfr_{nb}).
    \label{eq:pwc_wave_func}
    \ee
Here $\bfr_{na}$ and $\bfr_{nb}$ are the coordinates of the two paired electrons localized near WC lattice site $\bfR_n$, and $\Phi$ and $\varphi$ are the COM and relative wave functions of a pair, respectively. For the COM wave function we take a Gaussian with a variational width $\sigma_p$, while the relative wave function is determined by solving the two-electron pairing problem in a WC unit cell. In this approximation we include exchange effects within a pair (the spatial pair wave function should be fully antisymmetrized together the spin part of the wave function), but neglect exchange between pairs.

As we demonstrate explicitly below, using the Hartree-type wave functions \eqref{eq:wc_wave_func} and \eqref{eq:pwc_wave_func} yields  results identical to those of the second-quantized approach in the no-intercell-overlap approximation presented in the previous section. More comprehensive recent discussions of first-quantized wave functions in bands with nontrivial quantum geometry can be found in e.g., Refs.~\cite{tan2025,tan2025ideal}.

\subsection{Variational energy of the monatomic WC}

Using the wave function \eqref{eq:wc_wave_func}, the kinetic energy is readily found to to be  $\langle \Psi |  T  | \Psi \rangle = N \hbar ^2 / 4m^{*}\sigma^{2}$, in agreement with Eq.~\eqref{eq:t_wc}. The potential energy is
\begin{align}
    &\left \langle \Psi |  V | \Psi \right\rangle =\frac{N}{2}\sum_{\bfR_n\neq0}\int \dd^2\bfr_{1} \dd^2 \bfr_{2}\dd^2\bfr'_{1} \dd^2 \bfr'_{2}  ~ \Phi^{*}\left(\bfr_{1}\right)\Phi^{*}\left(\bfr_{2}-\bfR_n\right)\Phi\left(\bfr_{1}^{\prime}\right)\Phi\left(\bfr_{2}^{\prime}-\bfR_n\right)\Bra{\bfr_{1},\bfr_{2}}\hat{V}\Ket{\bfr_{1}^{\prime},\bfr_{2}^{\prime}}.
\end{align}
Plugging in the interaction matrix elements \eqref{eq:Vrrp} and integrating over the primed coordinates gives
    \be
    \begin{aligned}
    \left \langle \Psi |  V | \Psi \right\rangle  &= \frac{N}{2A}\sum_{\bfR_n\neq0} \sum_{\bfq \neq 0} v_\bfq e^{-\frac{\Omega}{2}q^{2}}e^{-i\bfq\cdot\bfR_n} \\
    &\qquad \times \int \dd^2\bfr_{1} \dd^2 \bfr_{2}  ~\Phi^{*}\left(\bfr_{1}\right)\Phi^{*}\left(\bfr_{2}\right)\Phi\left(\bfr_{1}+\frac{\left(\bfq\times\boldsymbol{\Omega}\right)}{2}\right)\Phi\left(\bfr_{2}-\frac{\left(\bfq\times\boldsymbol{\Omega}\right)}{2}\right)e^{i\bfq\cdot\left(\bfr_{1}-\bfr_{2}\right)}.
    \end{aligned}
    \ee
The integrals can be directly computed using the Gaussian form of the variational wave function, giving
\begin{align}
I_\pm\left(\bfq,\Omega\right)=\int \dd^2\bfr ~ e^{\pm i\bfq\cdot\bfr}\Phi^{*}\left(\bfr\right)\Phi\left(\bfr\pm\frac{\left(\bfq\times\boldsymbol{\Omega}\right)}{2}\right)=\exp\left\{ -\frac{q^{2}\sigma^{2}}{2}\left(1+\frac{\Omega^{2}}{16\sigma^{4}}\right)\right\}.
\end{align}
This gives the potential energy
\begin{align}
    &\left \langle \Psi |  V | \Psi \right\rangle =\frac{N}{2}\sum_{\bfR_n\neq0}\int \frac{\dd^2\bfq}{(2\pi)^2} ~ v_\bfq e^{-\tilde{\sigma}^{2}q^{2}}e^{-i\bfq\cdot\bfR_n}=\frac{Ne^{2}}{2}\sum_{\bfR_n\neq0}\frac{\sqrt{\pi}}{2\tilde{\sigma}}e^{ -\frac{R^2_{n}}{8\tilde{\sigma}^{2}}}I_{0}\left(\frac{R^{2}_n}{8\tilde{\sigma}^{2}}\right),
\end{align}
in agreement with Eq.~\eqref{eq:V_WC_I0} obtained using the second-quantized approach.

\subsection{Variational energy of the paired WC}

Taking the pair state wavefunction ansatz \eqref{eq:pwc_wave_func}, we proceed to the calculation of the paired crystal energy. The kinetic energy factorizes as a sum of the kinetic energies of the COM and relative coordinates of the pairs.
\be
    \langle \Psi_p | T | \Psi_p \rangle = N_{p}\frac{\hbar^{2}}{4M\sigma^{2}_{p}}+N_p \int \dd^{2}\bfr  ~ \varphi^{*}\left(\bfr\right)\frac{\hat{\bfp}^{2}}{2\mu}\varphi\left(\bfr\right), 
\ee
in agreement with Eq.~\eqref{eq:T_PWC}.

The calculation of potential energy here is more involved and must account for both the intra- and intercell Coulomb interactions. We write it as
\begin{align}
    & \langle \Psi_p |  V | \Psi_p \rangle = N_{p}v_{\text{intra}}+\frac{N_{p}}{2}v_{\text{inter}}.
\end{align}

\subsubsection{Intracell interaction}

For the intracell interaction, we have the expression
\begin{align}
    &v_{\text{intra}}=\int \dd^2\bfr_{1} \dd^2 \bfr_{2}\dd^2\bfr'_{1} \dd^2 \bfr'_{2} ~ \psi^{*}\left(\bfr_{1},\bfr_{2}\right)\psi\left(\bfr_{1}^{\prime},\bfr_{2}^{\prime}\right)\Bra{\bfr_{1},\bfr_{2}}\hat{V}\Ket{\bfr_{1}^{\prime},\bfr_{2}^{\prime}},
\end{align}
where $\psi(\bfr_1,\bfr_2) = \Phi\left( (\bfr_1 + \bfr_2)/2\right) \varphi(\bfr_1 - \bfr_2)$. 
Using the matrix elements \eqref{eq:Vrrp} and performing the integrals over primed coordinates, we obtain
\begin{align} v_{\text{intra}} &=\int \frac{\dd^2\bfq}{(2\pi)^2} ~ v_\bfq e^{-\frac{\Omega}{2}q^{2}}\int \dd^2\bfr_{1} \dd^2 \bfr_{2} ~ \psi^{*}\left(\bfr_{1},\bfr_{2}\right)\psi\left(\bfr_{1}+\frac{\left(\bfq\times\boldsymbol{\Omega}\right)}{2},\bfr_{2}-\frac{\left(\bfq\times\boldsymbol{\Omega}\right)}{2}\right)e^{i\bfq\cdot\left(\bfr_{1}-\bfr_{2}\right)},\\
&=\int \frac{\dd^2\bfq}{(2\pi)^2} ~ v_\bfq e^{-\frac{\Omega}{2}q^{2}}\int \dd^2\bfr ~  \varphi^{*}\left(\bfr\right)\varphi\left(\bfr+\left(\bfq\times\boldsymbol{\Omega}\right)\right)e^{i\bfq\cdot\bfr}. 
\end{align}
Fourier transforming $\varphi(\bfr) \to \tilde\varphi(\bfk)$ gives the intracell interaction energy in Eq.~\eqref{eq:Vintra_PWC}.

\subsubsection{Intercell interaction}

Now we compute the potential energy of the Coulomb interaction between pairs of electrons. Denote the particles in one unit cell by $1$ and $2$ and the particles in another unit cell by $1^{\prime}$ and $2^{\prime}$. Then the contribution to the potential energy coming from the interaction between the cells is given by
\begin{align}
v_{\text{inter}}&=v_{11^{\prime}}+v_{12^{\prime}}+v_{21^{\prime}}+v_{22^{\prime}} \\
&=\sum_{\bfR_n\neq0}\int_{\left\{ \bfr_{i}\right\} ,\left\{ \bfb_{i}\right\} }\psi^{*}\left(\bfr_{1},\bfr_{2}\right)\psi^{*}\left(\bfr_{1}^{\prime}-\bfR_n,\bfr_{2}^{\prime}-\bfR_n\right)\psi\left(\bfb_{1},\bfb_{2}\right)\psi\left(\bfb_{1}^{\prime}-\bfR_n,\bfb_{2}^{\prime}-\bfR_n\right) \nonumber \\
    &\qquad \times\left( \Bra{\bfr_{1}\bfr_{1}^{\prime}\bfr_{2}\bfr_{2}^{\prime}}\hat{V}_{11^{\prime}}\Ket{\bfb_{1}\bfb_{1}^{\prime}\bfb_{2}\bfb_{2}^{\prime}}+\Bra{\bfr_{1}\bfr_{2}^{\prime}\bfr_{2}\bfr_{1}^{\prime}}\hat{V}_{12^{\prime}}\Ket{\bfb_{1}\bfb_{2}^{\prime}\bfb_{2}\bfb_{1}^{\prime}} \right. \nonumber \\
    &\qquad ~~~~~  \left. +\Bra{\bfr_{2}\bfr_{1}^{\prime}\bfr_{1}\bfr_{2}^{\prime}}\hat{V}_{21^{\prime}}\Ket{\bfb_{2}\bfb_{1}^{\prime}\bfb_{1}\bfb_{2}^{\prime}}+\Bra{\bfr_{2}\bfr_{2}^{\prime}\bfr_{1}\bfr_{1}^{\prime}}\hat{V}_{22^{\prime}}\Ket{\bfb_{2}\bfb_{2}^{\prime}\bfb_{1}\bfb_{1}^{\prime}}\right) ,
\end{align}
where $\left\{ \bfr_{i}\right\}=\{\bfr_{1},\bfr_{2},\bfr_{1}^{\prime},\bfr_{2}^{\prime}\}$ and $\left\{ \bfb_{i}\right\} = \{\bfb_{1},\bfb_{2},\bfb_{1}^{\prime},\bfb_{2}^{\prime}\}$.
Using the fact that the electrons in a pair are indistinguishable, we recognize that all four terms in the sum above are equal. We thus focus on one representative, for example, $v_{12^\prime}$, such that
$v_{\text{inter}}=4v_{12^\prime}$.

The four-body matrix element of the interaction for this term has the form
\begin{equation}
\begin{aligned}
&\Bra{\bfr_{1}\bfr_{2}^{\prime}\bfr_{2}\bfr_{1}^{\prime}}\hat{V}_{12^{\prime}}\Ket{\bfb_{1}\bfb_{2}^{\prime}\bfb_{2}\bfb_{1}^{\prime}}=\Bra{\bfr_{1}\bfr_{2}^{\prime}}\hat{V}_{12^{\prime}}\Ket{\bfb_{1}\bfb_{2}^{\prime}}\delta\left(\bfr_{2}-\bfb_{2}\right)\delta\left(\bfr_{1}^{\prime}-\bfb_{1}^{\prime}\right).
\end{aligned}
\end{equation}
Putting in the matrix elements \eqref{eq:Vrrp} and performing the integrals using the delta functions, we arrive at
\begin{equation}
\begin{aligned}
v_\text{inter} &= \frac{1}{A}\sum_{\bfR_{n}\neq0}\sum_{\bfq \neq 0}  4v_\bfq e^{-\frac{\Omega}{2}q^{2}}e^{-i\bfq\cdot\bfR_{n}}\\
&\qquad \times\int \dd^2\bfR \dd^2\bfR^{\prime} ~ e^{i\bfq\cdot\left(\bfR-\bfR^{\prime}\right)}\Phi^{*}\left(\bfR\right)\Phi^{*}\left(\bfR^{\prime}\right)\Phi\left(\bfR+\frac{\left(\bfq\times\boldsymbol{\Omega}\right)}{4}\right)\Phi\left(\bfR^{\prime}-\frac{\left(\bfq\times\boldsymbol{\Omega}\right)}{4}\right) \\
&\qquad \times \int \dd^2\bfr \dd^2\bfr^{\prime} ~ e^{i\bfq\cdot\frac{\bfr+\bfr^{\prime}}{2}}\varphi^{*}\left(\bfr\right)\varphi^{*}\left(\bfr^{\prime}\right)\varphi\left(\bfr+\frac{\left(\bfq\times\boldsymbol{\Omega}\right)}{2}\right)\varphi\left(\bfr^{\prime}+\frac{\left(\bfq\times\boldsymbol{\Omega}\right)}{2}\right).
\end{aligned}
\end{equation}
Fourier transforming $\Phi(\bfr) \to \tilde\Phi(\bfq)$ and $\varphi(\bfr) \to \tilde\varphi(\bfk)$ we recover precisely the result \eqref{eq:v_inter} obtained in the second-quantization approach.

\section{Quantum mechanics in a topological band}

In this section, we briefly summarize the algorithm to obtain the first-quantized, wave function description of a particle in a band with nontrivial quantum geometry determined by the form factor \eqref{eq:form_factor}.

To account for both the Berry curvature and quantum metric of the band, one needs to perform the real-space analog of the Peierls substitution in the potential 
\be
    \hat\bfr\to \hat\bfr+\hat{\mathcal{A}}_{\bfk}
\ee
and dress the potential with the quantum metric factor
\begin{align}
    &v\left(\bfr\right)=\int\frac{\dd^{2}\bfq}{(2\pi)^2} {\left(2\pi\right)^{2}} ~ v_\bfq e^{i\bfq\cdot\bfr}\to v_{\Omega}\left(\bfr\right)=\int\frac{\dd^{2}\bfq}{\left(2\pi\right)^{2}} ~ v_\bfq e^{-\frac{\Omega}{4}q^{2}}e^{i\bfq\cdot\bfr}.
\end{align}

As an example, consider the Coulomb potential in 2D, $v_\bfq = 2\pi e^2/q$. A straightforward calculation gives
\be
    v\left(\bfr\right)=\frac{e^{2}}{r}\to v_{\Omega}\left(\bfr\right)=e^{2}\int_{0}^{\infty} \dd q  ~ J_{0}\left(qr\right)e^{-\frac{\Omega}{4}q^{2}}=e^{2}\sqrt{\frac{\pi}{\Omega}} e^{-\frac{r^{2}}{2\Omega}}  I_{0}\left(\frac{r^{2}}{2\Omega}\right).
\ee
For $r \gg \sqrt \Omega$, $v_\Omega$ tends to the usual Coulomb. However, at short distances $r \ll \sqrt\Omega$, the effective ``size" of the particle induced by quantum geometry cuts off the divergence in the Coulomb potential.  

For the harmonic potential, the action of the derivative on the exponential factor $\exp(-\Omega q^2/4)$ produces a shift
\begin{align}
    & v_{\text{harm}}\left(\bfr\right)=\frac{K\bfr^{2}}{2}\to v^{\Omega}_{\text{harm}}\left(\bfr\right)=\frac{K\bfr^{2}}{2}+\frac{K\Omega}{2}.
\end{align}

The single-particle behavior in a band with dispersion $\epsilon\left(\bfk\right)$ and uniform Berry curvature $\Omega$ with form factors \eqref{eq:form_factor} subjected to a potential $V(\bfr)$ is thus governed by the Hamiltonian
\begin{align}
    & \hat{H}_{\Omega}= \epsilon(\hat{\bfk})+V_{\Omega}(\hat{\bfr}+\hat{\mathcal{A}}_\bfk).
\end{align}

\end{document}